\documentclass{ws-ijmpa}
\usepackage{physics}
\usepackage{graphicx} 
\usepackage{array} 
    \newcounter{rowcntr}[table]
    \newcolumntype{N}{>{\refstepcounter{rowcntr}\alph{rowcntr}}c} 
    \AtBeginEnvironment{tabular}{\setcounter{rowcntr}{0}} 
\usepackage[utf8]{inputenc}
\usepackage[T1]{fontenc}
\usepackage{soul}
\usepackage{orcidlink}

\newcommand{\ang}[1]{\langle#1\rangle}
\newcommand{\ud}{\mathrm{d}}
\newcommand{\ue}{\mathrm{e}}
\newcommand{\ui}{\mathrm{i}}

\begin{document}
\title{An Upper Bound for the Mass of Microscopic Clocks}
\author{Bruno Arderucio Costa\footnote{bcosta@troy.edu} and Yafet E. Sanchez Sanchez\footnote{sanchezy@troy.edu}}
\address{Center for Relativity and Cosmology, Troy University,\\ Troy, Alabama, 36082, USA}
\date{\today}
\maketitle
\begin{abstract}
    According to general relativity, clocks are the basic measuring devices needed to probe spacetime geometry. However, it is generally accepted that the mass of clocks capable of measuring small time intervals must be bounded from below. In this article, we consider two gravitationally induced phenomena: first, the extent to which such a mass disturbs the geometry that the clocks intended to probe; second, the magnitude of the gravitational self-interaction. We adopt the semiclassical coupling between gravity and quantum matter in the non-relativistic regime to obtain \emph{upper} bounds on the mass of the clocks for a given time resolution and running time.
\end{abstract}

\section{Introduction}
Clocks are the standard measuring device in relativistic physics. If a set of clocks can be trusted to measure proper times accurately, it can be used, in principle, to measure all local observables~\cite{Wigner1957,Matsas2024}, in particular, the spacetime metric~\cite{geroch1981}. If clocks could be made arbitrarily light, these devices could serve as ideal instruments, extracting information from spacetime without disturbing it. Unfortunately, quantum mechanics establishes a minimum mass for a clock given its accuracy and running time~\cite{Salecker1958,Schumacher2010}. This poses an immediate conundrum: if one pursues sufficiently small clocks to extract information about the local geometry, accurate clocks will hence require large energy concentrations, making their gravitational back-reaction effects significant. Worse, because Einstein's equations are nonlinear, not even the most skillful clockmaker can account for the gravitational corrections in a generic spacetime. Rather, she can only do so with prior knowledge of the metric. Her clocks are thus unreliable for probing the spacetime metric.

Although gravity-induced limitations on time measurement are typically small, they are universal in nature. Much like the \emph{Gedankenexperiment} known as Heisenberg's microscope foreshadowed the uncertainty relations, the $\hbar$-, $c$- and $G$-dependent restrictions on time measurements may hint at a possible tenet from a future theory of quantum gravity.

In this article, we study the gravitational effects on quantum clocks. We adopt the framework of semiclassical gravity, in which the geometry responds to the expectation values of the stress--energy tensor. Since our primary interest is to determine how ambitious one can be in measuring time before the onset of gravitational effects, we shall keep all results to the first order in Newton's constant $G$. {For a given running time $T$ and accuracy $\tau$, we obtain two upper bounds for the mass of the clocks that scale differently with $T$ and $\tau$: the first comes from the failure of the coordinate time $t$ to represent the clock's proper time, and the second comes from the gravity-mediated self-interaction in the non-relativistic limit of the semiclassical Einstein equations, which is the Newton--Schrödinger equation~\cite{Bahrami2014}.}

\section{The Clock}
If one could neglect gravity, one could model a clock, as Salecker and Wigner~\cite[Section 3, Example b)]{Salecker1958} did, as a quantum simple harmonic oscillator described by the Hamiltonian
\begin{equation}
    H=\frac{1}{2}m\dot{\vec r}^2+\frac{1}{2}m\omega_0^2 \vec r^2,
    \label{hamiltonian}
\end{equation}
representing a spherically symmetric well of natural frequency $\omega_0$ in the energy-superposition state $|\psi_0\rangle=\frac{1}{\sqrt2}(|0\rangle+|N\rangle)$, where $|k\rangle$ denotes the spherically symmetric energy eigenstate of energy $E_k=\left(2k+\frac{3}{2}\right)\hbar\omega_0$. The probability of measuring the oscillator in its initial state $|\psi_0\rangle$ after a time $t>0$ is $p(t)=\cos^2(\omega_{0N}t/2)$ for $\omega_{0N}=\frac{E_N-E_0}{\hbar}$.

An experimenter {wishes to measure the coincidence rate of the oscillator to measure the elapsed time. As a prior $P(t)$, he adopts a uniform distribution in the interval $0\leq t\leq T$. The constant $T$ will be referred to as the running time. Because $p(t)$ is periodic, a running time $T\leq\frac{\pi}{2\omega_{0N}}$ avoids creating a periodic ambiguity in the elapsed time. He then uses} a collection of $\mathcal N$ identical oscillators and, after an unknown time $t$, count how many oscillators were found back in the original state. This fraction can be used, via Bayes' theorem, to improve the probability that the elapsed time falls within a certain range. For example, after $n$ oscillators were found in the initial state, the prior gets updated to the posterior probability distribution
\begin{equation}
    P_\text{new}(t)\propto\frac{2\Gamma(\mathcal N+1)p(t)^n(1-p(t))^{\mathcal N-n}}{\Gamma\left(n+\frac{1}{2}\right)\Gamma\left(\mathcal N-n+\frac{1}{2}\right)}P(t),
    \label{update}
\end{equation}
where $p(t)=|\langle\psi(t)|\psi_0\rangle|^2$. The precise proportionality constant depends on the running time $T$ of the clocks but not on $\mathcal N$.

    \begin{figure*}
        \centering
        \includegraphics[width=0.32\linewidth]{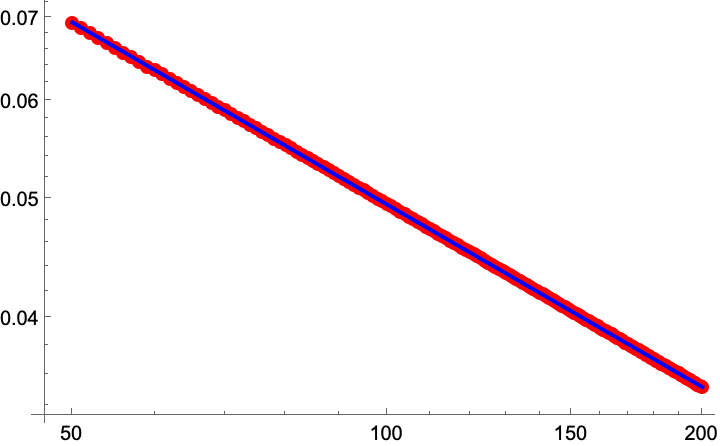}
        \includegraphics[width=.32\linewidth]{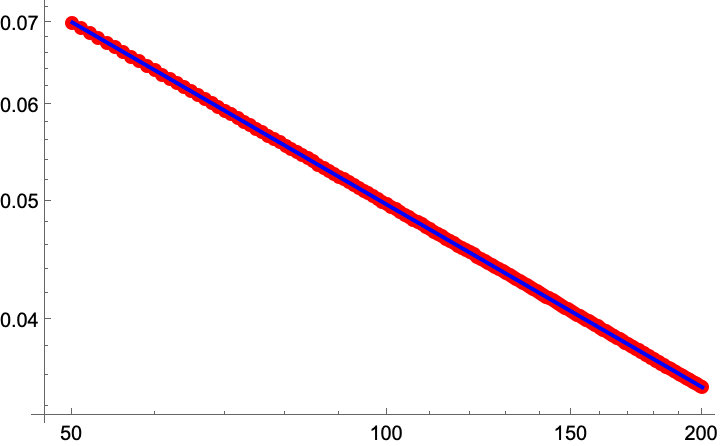}
        \includegraphics[width=.32\linewidth]{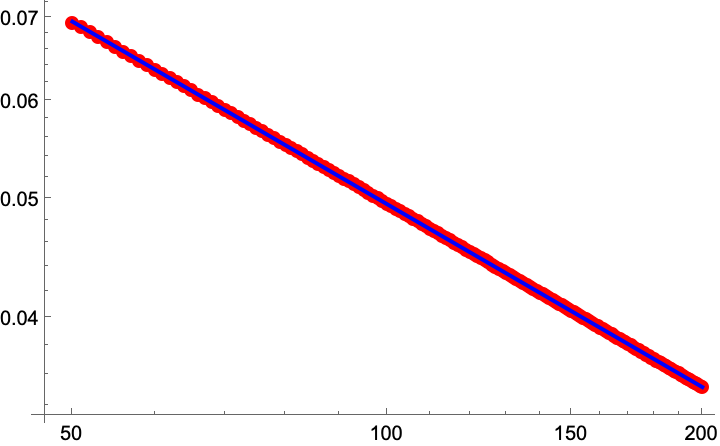}
        \caption{The variance $\tau$ for the distribution~\eqref{update} behaves approximately as the same power law~\eqref{nparticletau} for large $\mathcal N$ for three ratios $n/\mathcal N$: 10\% (left), 50\% (center), and 90\% (right). The calculated points (for $50\leq\mathcal N\leq200$) are in red, and the fitted curve is in blue. The fitted values of $\gamma$ were 0.490, 0.495, and 0.491, respectively. The largest difference in the value of $\gamma$ for a pair of ratios $n/\mathcal N$ is around 1\%. Fits for larger values of $\mathcal N$ bring $\gamma$ closer to \textonehalf.}
        \label{fits}
    \end{figure*}

If one wants to improve the time resolution of the clock, one can choose a large value for $\mathcal N$. We can evaluate how much the standard deviation reduces with an increment in $\mathcal N$ by fixing a ratio $f\equiv n/\mathcal N$ in Eq.~\eqref{update} and computing the posterior standard deviation for various values of $\mathcal N$. For large $\mathcal N$, expression~\eqref{update} simplifies to
\begin{equation}
P_\text{new}(t)\propto\sqrt{\frac{2\mathcal N}{\pi}}\frac{p(t)^{f\mathcal N}(1-p(t))^{(1-f)\mathcal N}}{f^{f\mathcal N}(1-f)^{(1-f)\mathcal N}}P(t).
    \label{updatelarge}
\end{equation}

For later use, if, as in the quantum harmonic oscillator in the state $|\psi_0\rangle$, $p(t)=\cos^2(\Omega t),\ (0\leq t\leq \frac{\pi}{2\Omega})$, the standard deviation for one clock is $\tau_1=20.56\%\frac{\pi}{2\Omega}$. And, for large $\mathcal N$, the standard deviation of $\tau$ of the posterior $P_\text{new}(t)$ is approximated by a power law
\begin{equation}
    \tau\sim K\mathcal N^{-\gamma},
    \label{nparticletau}
\end{equation}
for $\gamma\approx\frac{1}{2}$ for all ratios $n/\mathcal N$ (Figure~\ref{fits}). Here, the fitted value for $K$ is greater than $\tau_1$, but of the same order of magnitude. Henceforth, we adopt $\tau$ as a measure of the clock's time resolution: The higher the accuracy, the smaller the value of $\tau$.

However, increasing the number of oscillators is not a foolproof improvement. Clocks can only measure the same time interval if they travel along the same worldline. Consequently, an improvement in statistics necessarily comes at the cost of a more significant gravitational disturbance in the metric one wishes to probe.

\section{Semiclassical coupling}
To account for the clock's backreaction on the metric, one needs a model of how quantum matter couples to gravity. One such model is the semiclassical coupling, in which the Einstein tensor responds to the renormalized expectation value of the stress--energy tensor of a quantum field in a given state $\sigma$. 
That is,  $G_{ab}=\frac{8\pi G}{c^4}\langle T_{ab}\rangle_{\sigma,\text{ren}}$. Since we wish to model several clocks in the same state, we must consider a bosonic field, which, for simplicity, we take to be scalar. If $\sigma$ represents a state with $\mathcal N$ particles in the same state (emulating $\mathcal N$ identical clocks), it can be written in the form $\frac{1}{\sqrt{\mathcal N!}}(a_k^\dagger)^\mathcal N|0\rangle$, where $|0\rangle$ is the reference state for which $\langle T_{ab}\rangle_{|0\rangle,\text{ren}}=0$ and $a_k^\dagger$ is a creation operator for a mode $u_k$, i.e., $a_k^\dagger=(\Phi,u_k)$, where $\Phi$ is the field operator, and the brackets denote the Klein--Gordon product. Because the stress--energy tensor is evaluated as $\mathcal T_{ab}(\Phi,\Phi)$ for a tensor $\mathcal T_{ab}(u_i,u_j)$ that is a bilinear function of its arguments $u_i$ and $u_j$, writing the field operator as $\Phi=\sum_i(u_i a_i+\bar u_i a_i^\dagger)$, one has 
\begin{equation*}
    \langle T_{ab}\rangle_\sigma=\frac{1}{\mathcal N!}\sum_{i,j}\mathcal T_{ab}(u_i,\bar u_j)\langle 0|a_k^\mathcal Na_i a_j^\dagger(a^\dagger_k)^\mathcal N|0\rangle+\mathcal T_{ab}(\bar u_i,u_j)\langle0|a_k^\mathcal Na_i^\dagger a_j(a^\dagger_k)^\mathcal N|0\rangle.
\end{equation*}

From the canonical commutation relations $[a_i,a_j^\dagger]=\delta_{ij}$, one can prove the identity $a_j(a_k^\dagger)^\mathcal N|0\rangle=\mathcal N (a_k^\dagger)^{\mathcal N-1}\delta_{jk}|0\rangle$ by induction. Applying it twice to the above expression,
\begin{equation}
    \langle T_{ab}\rangle_\sigma=\langle T_{ab}\rangle_0+\mathcal N\ 2\mathrm{Re}\ \mathcal T_{ab}(u_k,u_k).
    \label{nclockstate}
\end{equation}

Equation~\eqref{nclockstate} states that, unsurprisingly, the stress--energy tensor on an $\mathcal N$-particle state scales, after renormalization, with $\mathcal N$.

{
It is nontrivial to take the non-relativistic limit of the semiclassical equations, or even of quantum field theories~\cite{falcone2023nonrelativistic,Anastopoulos2014,Bahrami2014}. Nonetheless, under suitable approximations, it leads to the Newton--Schr{\"o}dinger equation. In the usual Minkowski quantization, a complete set of modes $\tilde u_k$ are taken to be $\tilde u_k\propto \ue^{\ui(\mathbf k\cdot\mathbf r-\omega_kt)}$ with $\omega^2=k^2c^2+m^2c^4/\hbar^2$. For the state $\sigma$ to represent a collection of non-relativistic excitations in Minkowski vacuum, the modes $u_i$ can only be built from $\tilde u_i$ for which $k^2\ll m^2c^2/\hbar^2$.
In this approximation,  the functions $$\Psi(\mathbf r_1,\ldots,\mathbf r_\mathcal N)=\langle0|\prod_{i=1}^\mathcal N \Phi(\mathbf r_i)|\sigma\rangle$$ form the Hilbert space of a non-relativistic quantum theory of $\mathcal N$ particles. Although any Hamiltonian compatible with the non-relativistic approximation must commute with the number operator (after all, all interaction energy scales must be much smaller than $mc^2$), one can still write the non-relativistic theory in the language of second quantization of a Schrödinger field $\psi$.

In this framework, together with the Newtonian approximation in general relativity, i.e., $g_{\mu\nu}=\eta_{\mu\nu}+h_{\mu\nu}$ with the only nonzero $h_{\mu\nu}$ being 
$h_{00}=2\frac{\phi}{c^2}$ and $\langle T_{00}\rangle_\sigma=mc^2\ang{\Psi|\psi(x)\psi^\dagger(x)|\Psi}$ being the only nonzero component of the stress--energy tensor, the equations of motion for the $\mathcal N$-particle wave function $\Psi(\mathbf r_1,\ldots,\mathbf r_\mathcal N)$ are the Newton--Schr{\"o}dinger equation 

\begin{equation} \label{nse}
\ui\hbar \pdv{\Psi}{t} = \left[ \underbrace{-\sum_{i=1}^{\cal{N}} \frac{\hbar^2}{2m_i} \nabla_i^2 +\sum_{i=1}^{\cal{N}} \omega_0r_i^2}_{H_0} + \sum_{i=1}^{\cal{N}} m_i \phi(\vb{r}_i, t) \right] \Psi,
\end{equation}
in agreement with the results of Ref.~\cite[(Eq. 12)]{Bahrami2014}. Here,
\( \phi(\vb{r}, t) \) is the gravitational potential determined by the Newtonian approximation of the semiclassical Einstein equations, which leads to
\begin{equation} \label{Phi}
\phi(\vb{r}, t) = -G \int \frac{\rho(\vb{r}', t)}{\abs{\vb{r} - \vb{r}'}} \, \ud^3 r',
\end{equation}
where the mass density \( \rho(\vb{r}, t) \) is given by
\begin{equation} \label{rho}
\rho(\vb{r}, t)=\sum_{i=1}^{\cal{N}} m_i \int \abs{\Psi(\vb{r}_1, \dots, \vb{r}_{\cal{N}}, t)}^2 \delta(\vb{r} - \vb{r}_i) \, \ud^3 r_1 \dots \ud^3 r_\mathcal N.
\end{equation}
}

Equation \eqref{nse} describes the dynamics of the clock when its self-gravity has been taken into account. Notice that $H_0$ is ${\cal{N}}$ copies of the Hamiltonian~\eqref{hamiltonian} describing ${\cal{N}}$ quantum harmonic oscillators. 

For simplicity, we solve Eq.\eqref{nse} to the first order in $G$, i.e., we put $\Psi(t)=\Psi^0(t)+G\Psi^1(t)$ and $\phi=\phi^0+G\phi^1$ and solve for $\Psi^1(t)$ and $\phi^1$ (see details in Ref.~\cite[Chapter 12.5]{Ballentine2014}). 

To be precise, to order $0$, $\Psi^0$ is given by $\cal{N}$ independent quantum harmonic oscillators in the energy-superposition state $|\psi_0\rangle$ described above and the potential $\phi^0=0$. To the first order, $\phi^1$ is given by Eq.\eqref{Phi}, with $\rho$ given by $\rho^0$, the zero order in $\rho$, which in this case is given by ${\cal{N}}|\psi_0|^2$. Hence,
$$\phi^1=-{\cal{N}}m\int\frac{|\psi^0(r')|^2}{|\Vec{r}-\Vec{r'}|}\ud^3r'$$
and $\Psi^1$ satisfies 

\begin{equation}\label{nse1}
\ui\hbar \pdv{\Psi^1}{t} = \left[ \underbrace{-\sum_{i=1}^{\cal{N}} \frac{\hbar^2}{2m_i} \nabla_i^2 +\sum_{i=1}^{\cal{N}} \omega_0r_i^2}_{H_0} \right] \Psi^1+ \underbrace{\frac{1}{2}\sum_{i=1}^{\cal{N}} m_i \phi^1(\vb{r}_i, t) }_{\phi(\Psi^0)}\Psi^0.
\end{equation}

Inserting the ansatz 

\begin{equation}\label{ansatz}
  \Psi^1(t)=\sum_{{i_1,i_2,..i_{\cal{N}}}}C_{{i_1,i_2,..i_{\cal{N}}}}(t)\ue^{\frac{E_{{i_1,i_2,..i_{\cal{N}}}}t}{\hbar}}|{i_1,i_2,..i_{\cal{N}}}\rangle  ,
\end{equation}
where $|{i_1,i_2,..i_{\cal{N}}}\rangle$ is the state such that the $j$-th oscillator is in the energy eigenstate $i_j$, into Eq.\eqref{nse1} and using that $\Psi^0=|\psi_0\rangle\otimes|\psi_0\rangle\otimes\ldots\otimes|\psi_0\rangle$, we obtain a solution provided that

\begin{multline}
C_{i_1, \dots, i_{\cal N}}(t)= \frac{\ui m}{2\hbar} \int_0^t \exp\left( \frac{\ui}{\hbar} E_{i_1, \dots, i_{\cal N}} s \right) \times\\
\mel{i_1, \dots, i_{\cal N}}{\phi(\Psi^0)}{
\bigotimes_{j=1}^{\cal N} \left(\ue^{-\ui E_0 s/\hbar} \ket{0}_j + \ue^{-\ui E_N s/\hbar} \ket{1}_j \right)
} \, \mathrm{d}s.
\label{coefficients}
\end{multline}

Here, $E_j$ denotes the energy of the state $ \ket{j}$ of one oscillator. More generally, the energy $ E_{i_1, i_2, \dots, i_{\cal N}} $ corresponds to the eigenvalue of the unperturbed Hamiltonian $H_0$ associated with the product eigenstate $ \ket{i_1, i_2, \dots, i_{\cal N}} $, where the $ j $-th particle is in the one-particle eigenstate labeled by $i_j$. Because $\Psi^0$ is factorizable, $\phi(\Psi^0)$ takes the form ${\phi_1}\otimes I\otimes\dots+I\otimes {\phi_1}\otimes I\otimes\ldots+\ldots$. Hence, {the matrix element in Eq.~\eqref{coefficients} is}

\begin{multline}
\sum_{k=1}^{\cal N}
     \Bigl[
        \ue^{-\ui E_0 s/\hbar}\,\delta_{i_k,0}\,
        \mel{0}{{\phi_1}}{0}
       +\ue^{-\ui E_N s/\hbar}\,\delta_{i_k,1}\,
        \mel{N}{{\phi_1}}{N}+\ue^{-\ui E_N s/\hbar}\,\delta_{i_k,0}\,
        \mel{0}{{\phi_1}}{N}\\
       +\ue^{-\ui E_0 s/\hbar}\,\delta_{i_k,1}\,
        \mel{1}{{\phi_1}}{0}
     \Bigr]\!
     \prod_{\substack{j=1\\ j\neq k}}^{\cal N}
        \ue^{-\ui E_{i_j}s/\hbar}.
\label{eq:Phi1_matrix_element}
\end{multline}

\section{An Upper Bound for the Mass of the Clock}
In order for the model to fall within its range of applicability, that is, for the clock to function as a reliable timekeeper, we must remain within the domain of validity of our perturbative expansion.

The first order perturbation of the state $\Psi$ will be reliable in an interval of time $t\in [0,T]$ provided that  
\begin{equation}
    t\mel{i_1,\dots,i_{\cal N}}{\phi^{1}}
      {\bigotimes_{j=1}^{\cal N}
         \left(e^{-iE_0 s/\hbar}\ket{0}_j
               +e^{-iE_N s/\hbar}\ket{N}_j\right)}\ll1. 
\end{equation}


Choosing the first excited state ($N=1$), we obtain
\begin{equation}\label{boundNS}
   {\cal{N}}^2 G\omega^{-\frac{1}{2}} m^{\frac{5}{2}}\hbar^{-\frac{3}{2}}\ll1.
\end{equation}
 From Eq.~\ref{nparticletau}, $\frac{1}{\omega}\sim T\sim 5\tau \mathcal N^{\gamma}$. Therefore, we can rewrite the bound as 

\begin{equation}
   {\cal{N}}^2 G m^{\frac{5}{2}}T^{\frac{1}{2}}\hbar^{-\frac{3}{2}}\ll1.
   \label{nscond}
\end{equation}

 Now, in the non-relativistic, weak-field limit of general relativity, the metric \( g_{\mu\nu} \) can be written as $g_{\mu\nu} = \eta_{\mu\nu} + h_{\mu\nu}$, where \( |h_{\mu\nu}| \ll 1 \)~\cite{misner1973gravitation}.

Since $h_{00}=2 \frac{\phi}{c^2}$, we have that the metric is correctly described in perturbation theory as long as

\begin{equation}
   {\cal{N}}G m^{\frac{3}{2}}T^{-\frac{1}{2}}\hbar^{-\frac{1}{2}}c^{-2}\ll1.
   \label{grcond}
\end{equation}

{Combining Eqs.~\eqref{nscond} and~\eqref{grcond} and eliminating $\mathcal N$ from Eq.~\eqref{nparticletau},} our main result from the model is that a system of non-relativistic clocks can only probe proper times within the weak-field approximation reliably if the mass of each clock, $M_\text{clock}={\cal{N}}m$ respects

\begin{equation}
  M_{clock}\ll T^{\frac{1+\gamma}{3\gamma}}\hbar^{\frac13}c^{\frac43}\tau^{-\frac{1}{3\gamma}}G^{-\frac23}.
    \label{main}
\end{equation}

 The estimate itself is sound as long as the perturbation of the wave function approach is valid, which happens when

\begin{equation}
M_{clock}\ll \hbar^\frac{3}{5}G^{-\frac{2}{5}}T^\frac{1-\gamma}{5\gamma}\tau^{-\frac{1}{5\gamma}}.
    \label{condition}
\end{equation}

The bound~\eqref{main} decreases faster than the one in Eq.~\eqref{condition}. Adopting $\gamma=1/2$ (see Eq.\eqref{nparticletau}) as an approximation for large $\mathcal N$, condition~\eqref{main} shows a dependence of $T\tau^{-\frac{2}3}$, and Eq.~\eqref{condition} scales as {$T^{\frac{1}{5}}\tau^{-\frac{2}{5}}$.} In SI units, these conditions are
\begin{equation}
    M_\text{clock}\ll(4.83\cdot10^{-17}\mathrm{kg})\left(\frac{T}{\mathrm{sec}}\right)^\frac15\left(\frac{\tau}{\mathrm{sec}}\right)^{-\frac25},(5.76\cdot10^6\mathrm{kg})\left(\frac{T}{\mathrm{sec}}\right)\left(\frac{\tau}{\mathrm{sec}}\right)^{-\frac23}
    \label{bounds}
\end{equation}

{{We emphasize that a failure of compliance to Eq.~\eqref{condition} means more than the need for a second-order perturbative correction. Rather, it exposes that, in an appropriate semiclassical limit, the self-gravitational effect must be accounted for when the clock is manufactured, otherwise its state evolution would deviate considerably from the zeroth-order solution.}}

The bound on the mass of a single oscillator that comes from the weak-field approximation of general relativity decreases for {shorter} running times and higher accuracies. For a fixed running time, $T$, improving the time resolution (i.e., decreasing $\tau$) leads to a more stringent constraint, lowering the allowed mass range. This follows from the fact that to achieve higher time resolution requires higher energies, and therefore a higher energy density enhances the gravitational backreaction, eventually invalidating the weak-field approximation.
By contrast, the corresponding bound derived from the first-order perturbative treatment of the wave equation exhibits the opposite trend for finer time resolutions: the allowed mass increases for smaller values of $\tau$. This reflects the energy-time uncertainty relation, according to which shorter interaction times permit larger fluctuations in energy, allowing larger masses without breaking the perturbative description. {For $T>t_\text{Planck}\sim10^{-44}\mathrm{s}$, the bound obtained from the perturbative approximation of the wave function becomes the dominant constraint.}

\section{Discussion}

 As Peres argued~\cite{Peres1979}, for clocks to be useful, they must be coupled to the system they are designed to probe, and the more accurate a clock is, the more it interferes with the said system. Since gravity couples to all matter, its impacts on time resolution are potentially universal.

From an operationalist point of view, the observation establishes a parallel between our analysis and the classic Heisenberg microscope, which, even before the uncertainty relations established the foundations of the modern quantum theory, suggested that one could not know the position and momentum of an electron simultaneously. If a photon is used to localize an electron, an increase in the precision must be realized by a reduction in the photon's wavelength, which, in turn, increases the disturbance in the momentum of the electron. In our model, clocks play the role of the photon, and proper times (or equivalently, the spacetime metric) play the role of the electron. Much like small wavelengths improve the measurement of the position at the expense of the momentum in Heisenberg's argument, an improvement of the time resolution generates a greater deviation of the probed metric.\\

\bibliographystyle{ws-ijmpa}
\bibliography{bibliography}
\end{document}